\documentclass[manuscript]{aastex}

\usepackage{hyperref}
\usepackage{url}
\usepackage{lscape}
\usepackage{longtable}
\usepackage{latexsym}
\usepackage{graphics}
\usepackage{graphicx}
\usepackage{enumitem}
\usepackage{amsmath}
\usepackage{mathptmx}
\usepackage{latexsym}
\usepackage{fix-cm}
\usepackage{bigstrut}
\usepackage{booktabs}

\begin{document}

\title{A new class of solutions of anisotropic charged distributions on pseudo-spheroidal spacetime}

\author{B. S. Ratanpal\altaffilmark{1}}
\affil{Department of Applied Mathematics, Faculty of Technology \& Engineering, The M. S. University of Baroda, Vadodara - 390 001, India}
\email{bharatratanpal@gmail.com}

\author{V. O. Thomas\altaffilmark{2}}
\affil{Department of Mathematics, Faculty of Science, The M. S. University of Baroda,\\ Vadodara - 390 002, India}
\email{votmsu@gmail.com}

\and

\author{D. M. Pandya\altaffilmark{3}}
\affil{Department of Mathematics \& Computer Science, Pandit Deendayal Petroleum University, Raisan, Gandhinagar - 382 007, India}
\email{dishantpandya777@gmail.com}

\begin{abstract}

In the present article a new class of exact solutions of Einstein's field equations for charged anisotropic distribution is obtained
on the background of pseudo-spheroidal spacetime characterized by the metric potential $g_{rr}=\frac{1+K\frac{r^{2}}{R^{2}}}{1+\frac{r^{2}}{R^{2}}}$,
where $K$ and $R$ are geometric parameters of the spacetime. The radial pressure $p_{r}$ and electric field intensity $E$ are
taken in the form $8\pi p_{r}=\frac{K-1}{R^{2}}\frac{\left(1-\frac{r^{2}}{R^{2}} \right)}{\left(1+K\frac{r^{2}}{R^{2}} \right)^{2}}$
and $E^{2}=\frac{\alpha(K-1)\frac{r^{2}}{R^{2}}}{R^{2}\left(1+K\frac{r^{2}}{R^{2}} \right)^{2}}$. The bounds of geometric parameter
$K$ and the parameter $\alpha$ appearing in the expression of $E^{2}$ are obtained by imposing the requirements for a physically acceptable model. 
It is found that the model is in good agreement with the observational data of number of compact stars like 
4U 1820-30, PSR J1903+327, 4U 1608-52, Vela X-1, PSR J1614-2230, Cen X-3 given by Gangopadhyay {\em{et al}} [Gangopadhyay T., Ray S., Li X-D., Dey J. and Dey M., {\it Mon. Not. R. Astron. Soc.} {\bf431} (2013) 3216]. When $ \alpha = 0, $ the model reduces to the uncharged anisotropic distribution given by Ratanpal {\em et al.} [Ratanpal B. S., Thomas V. O. and Pandya D. M., arXiv:1506.08512 [gr-qc](2015)]

\end{abstract}


\section{Introduction}
Mathematical model for generating superdense compact star models compatible with observational data has got wide
attention among researchers. A number of papers have been appeared in literature in the recent past along this direction
considering matter distribution incorporating charge \citep{Maurya11a,Maurya11b,Maurya11c,Pant12,Maurya15}. It has been suggested, as a result of theoretical investigations of \cite{Ruderman72} and \cite{Canuto74}, that matter may not be isotropic in high density regime of $10^{15}~gm/cm^{3}$. Hence it is pertinent to
construct charged distribution incorporating anisotropy in pressure. \cite{Bonner60, Bonner65} has shown
that a spherical distribution of matter can retain its equilibrium by counter balancing the gravitational force of attraction by
Coulombian force of repulsion due to the presence of charge. It was shown by \cite{Stettner73} that a spherical distribution
of uniform density accompanied by charge is more stable than distribution without charge. The study of charge distributions on 
spheroidal spacetimes have been carried out by \cite{Patel87}, \cite{Singh98}, \cite{Sharma01}, \cite{Gupta05}, \cite{Komathiraj07}. The spheroidal spacetime is found to accommodate superdense stars like neutron stars in both charged and uncharged cases. Study of strange stars and quark stars in the presence of electric charge have been done by \cite{Sharma06}, \cite{Mukherjee01}, \cite{Mukherjee02}. Recently charged fluid models have also been studied by  \cite{Maurya11a,Maurya11b,Maurya11c}, \cite{Pant12} \& \cite{Maurya15}.

In this paper, we have obtained a new class of solutions for charged fluid distribution on the background of pseudo spheroidal spacetime.
Particular choices for radial pressure $p_{r}$ and electric field intensity $E$ are taken so that the physical requirements and regularity conditions are not violated. The bounds for the geometric parameter $K$ and the parameter $\alpha$ associated with charge, are determined using various physical requirements that are expected to satisfy in its region of validity. It is found that these models can accommodate a number of pulsars like 4U 1820-30, PSR J1903+327, 4U 1608-52, Vela X-1, PSR J1614-2230, Cen X-3, given by \cite{Gangopadhyay13}. When $ \alpha = 0, $ the model reduces to the uncharged anisotropic distribution given by \cite{Ratanpal15}.

In section \ref{sec:2}, we have solved the field equations and in section \ref{sec:3}, we have obtained
the bounds for different parameters using physical acceptability and regularity conditions. In section \ref{sec:4}, 
We have displayed a variety of pulsars in agreement wit the charged pseudo-spheroidal model developed. In particular we have studied a model for
various physical conditions throughout the distribution and discussed the main results at the end of this section.

\section{Spacetime Metric}
\label{sec:2}
We shall take the interior spacetime metric representing charged anisotropic matter distribution as
\begin{equation}\label{IMetric1}
	ds^{2}=e^{\nu(r)}dt^{2}-\left(\frac{1+K\frac{r^{2}}{R^{2}}}{1+\frac{r^2}{R^{2}}} \right)dr^{2}-r^{2}\left(d\theta^{2}+\sin^{2}\theta d\phi^{2} \right),
\end{equation}
where $K$ and $R$ are geometric parameters and $K>1$. This spacetime, known as pseudo-spheroidal spacetime, has been studied by number
of researchers~\cite {Tikekar98,Tikekar99,Tikekar05,Thomas05,Thomas07,Paul11,Chattopadhyay10,Chattopadhyay12} have found that it can accommodate compact superdense stars.

Since the metric potential $g_{rr}$ is chosen apriori, the other metric potential $\nu\left(r \right)$ is to be determined by solving
the Einstein-Maxwell field equations
\begin{equation}\label{FE}
	R_{i}^{j}-\frac{1}{2}R\delta_{i}^{j}=8\pi\left(T_{i}^{j}+\pi_{i}^{j}+E_{i}^{j} \right),
\end{equation}
where,
\begin{equation}\label{Tij}
	T_{i}^{j}=\left(\rho+p \right)u_{i}u^{j}-p\delta_{i}^{j},
\end{equation}
\begin{equation}\label{piij}
	\pi_{i}^{j}=\sqrt{3}S\left[c_{i}c^{j}-\frac{1}{2}\left(u_{i}u^{j}-\delta_{i}^{j} \right) \right],
\end{equation}
and
\begin{equation}\label{Eij}
	E_{i}^{j}=\frac{1}{4\pi}\left(-F_{ik}F^{jk}+\frac{1}{4}F_{mn}F^{mn}\delta_{i}^{j} \right).
\end{equation}
Here $\rho$, $p$, $u_{i}$, $S$ and $c^{i}$, respectively, denote the proper density, fluid pressure, unit-four velocity, magnitude
of anisotropic tensor and a radial vector given by $\left(0, -e^{-\lambda/2}, 0, 0 \right)$. $F_{ij}$ denotes the anti-symmetric
electromagnetic field strength tensor defined by 
\begin{equation}\label{Fij}
	F_{ij}=\frac{\partial A_{j}}{\partial x_{i}}-\frac{\partial A_{i}}{\partial x_{j}},
\end{equation}
which satisfies the Maxwell equations
\begin{equation}\label{ME1}
	F_{ij,k}+F_{jk,i}+F_{ki,j}=0,
\end{equation}
and
\begin{equation}\label{ME2}
	\frac{\partial}{\partial x^{k}}\left(F^{ik}\sqrt{-g} \right)=4\pi\sqrt{-g}J^{i},
\end{equation}
where $g$ denotes the determinant of $g_{ij}$, $A_{i}=\left(\phi(r), 0, 0, 0 \right)$ is four-potential and
\begin{equation}\label{Ji}
	J^{i}=\sigma u^{i},
\end{equation}
is the four-current vector where $\sigma$ denotes the charge density.

The only non-vanishing components of $F_{ij}$ is $F_{01}=-F_{10}$. Here
\begin{equation}\label{F01}
	F_{01}=-\frac{e^{\frac{\nu+\lambda}{2}}}{r^{2}}\int_{0}^{r} 4\pi r^{2}\sigma e^{\lambda/2}dr,
\end{equation}
and the total charge inside a radius $r$ is given by 
\begin{equation}\label{qr}
	q(r)=4\pi\int_{0}^{r} \sigma r^{2}e^{\lambda/2}dr.
\end{equation}
The electric field intensity $E$ can be obtained from $E^{2}=-F_{01}F^{01}$, which subsequently reduces to
\begin{equation}\label{E}
	E=\frac{q(r)}{r^{2}}.
\end{equation}
The field equations given by (\ref{FE}) are now equivalent to the following set of the non-linear ODE's
\begin{equation}\label{FE1}
	\frac{1-e^{-\lambda}}{r^{2}}+\frac{e^{-\lambda}\lambda'}{r}=8\pi\rho+E^{2},	
\end{equation}
\begin{equation}\label{FE2}
	\frac{e^{-\lambda}-1}{r^{2}}+\frac{e^{-\lambda}\nu'}{r}=8\pi p_{r}-E^{2},
\end{equation}
\begin{equation}\label{FE3}
	e^{-\lambda}\left(\frac{\nu''}{2}+\frac{\nu'^{2}}{4}-\frac{\nu'\lambda'}{4}+\frac{\nu'-\lambda'}{2r} \right)=8\pi p_{\perp}+E^{2},
\end{equation}
where we have taken
\begin{equation}\label{pr1}
	p_{r}=p+\frac{2S}{\sqrt{3}},
\end{equation}
\begin{equation}\label{pp1}
	p_{\perp}=p-\frac{S}{\sqrt{3}}.
\end{equation}
Because $e^{\lambda}=\frac{1+K\frac{r^{2}}{R^{2}}}{1+\frac{r^{2}}{R^{2}}}$, the metric potential $\lambda$ is known function of $r$. The set of equations (\ref{FE1}) - (\ref{FE3}) are to
be solved for five unknowns $\nu$, $\rho$, $p_{r}$, $p_{\perp}$ and $E$. So we have two free variables for which suitable assumption
can be made. We shall assume the following expressions for $p_{r}$ and $E$.
\begin{equation}\label{pr2}
	8\pi p_{r}=\frac{K-1}{R^{2}}\frac{1-\frac{r^{2}}{R^{2}}}{\left(1+K\frac{r^{2}}{R^{2}} \right)^{2} },
\end{equation} 
\begin{equation}\label{E2}
	E^{2}=\frac{\alpha\left(K-1 \right)}{R^{2}}\frac{\frac{r^{2}}{R^{2}}}{\left(1+K\frac{r^{2}}{R^{2}} \right)}.
\end{equation}
It can be noticed from equation (\ref{pr2}) that $p_{r}$ vanishes at $r=R$ and hence we take the geometric parameter $R$
as the radius of distribution. Further $p_{r}\geq 0$ for all values of $r$ in the range $0\leq r\leq R$. It can also be noted that
$E^{2}$ is regular at $r=0$. On substituting the values of $p_{r}$ and $E^{2}$ in (\ref{FE2}) we obtain, after a lengthy 
calculation
\begin{equation}\label{enu}
	e^{\nu}=CR^{\frac{\left[K^{2}-(2+\alpha)K+\alpha+1 \right]}{K}}\left(1+K\frac{r^{2}}{R^{2}} \right)^{\left(\frac{K+\alpha+1}{2K} \right)}\left(1+\frac{r^{2}}{R^{2}} \right)^{\frac{K-\alpha-3}{2}},
\end{equation}
where $C$ is a constant of integration. Hence, the spacetime metric takes the explicit form
\begin{eqnarray}\label{IMetric2}
	ds^{2} & = & CR^{\frac{\left[K^{2}-(2+\alpha)K+\alpha+1 \right]}{K}}\left(1+K\frac{r^{2}}{R^{2}} \right)^{\left(\frac{K+\alpha+1}{2K} \right)}\left(1+\frac{r^{2}}{R^{2}} \right)^{\frac{K-\alpha-3}{2}}dt^{2}\\\nonumber
	       &   & -\left(\frac{1+K\frac{r^{2}}{R^{2}}}{1+\frac{r^{2}}{R^{2}}} \right)dr^{2}-r^{2}\left(d\theta^{2}+\sin^{2}\theta d\phi^{2}\right).
\end{eqnarray}
The constant of integration $C$ can be evaluated by matching the interior spacetime metric with Riessner-Nordstr{\"o}m metric
\begin{equation}\label{EMetric2}
	ds^{2}=\left(1-\frac{2m}{r}+\frac{q^{2}}{r^{2}} \right)dt^{2}-\left(1-\frac{2m}{r}+\frac{q^{2}}{r^{2}} \right)^{-1}dr^{2}-r^{2}\left(d\theta^{2}+\sin^{2}\theta d\phi^{2}\right),
\end{equation}
across the boundary $r=R$. This gives
\begin{equation}\label{M}
	M=\frac{R}{2}\frac{\left[K^{2}+\alpha(K-1)-1 \right]}{\left(1+K \right)^{2}},
\end{equation}
and
\begin{equation}\label{C}
	C=R^{\frac{-\left[K^2-(2+\alpha)K+\alpha+1 \right]}{K}}\left(1+K \right)^{-\left(\frac{3K+\alpha+1}{2K} \right)}2^{\left(\frac{\alpha-K+5}{2} \right)}.
\end{equation}
Here $M$ denotes the total mass of the charged anisotropic distribution.
\section{Physical Requirements and Bounds for Parameters}
\label{sec:3}
The gradient of radial pressure is obtained from equation (\ref{pr2}) in the form
\begin{equation}\label{dprdr}
	8\pi\frac{dp_{r}}{dr}=-\frac{2r(K-1)}{R^{4}}\frac{1+2K-K\frac{r^{2}}{R^{2}}}{\left(1+K\frac{r^{2}}{R^{2}} \right)^{3}}<0.
\end{equation}
It can be noticed from equation (\ref{dprdr}) that the radial pressure is decreasing function of $r$. 
Now, equation (\ref{FE1}) gives the density of the distribution as 
\begin{equation}\label{rho3}
	8\pi\rho=\left(\frac{K-1}{R^{2}}\right)\frac{3+(K-\alpha)\frac{r^{2}}{R^{2}}}{\left(1+K\frac{r^{2}}{R^{2}} \right)^{2}}.
\end{equation}
The conditon $\rho(r=0)>0$ is clearly satisfied and $\rho(r=R)>0$ gives the following inequality connecting $\alpha$ and $K$.
\begin{equation}\label{In1}
	0 \leq \alpha<3+K.
\end{equation}
Differentiating (\ref{rho3}) with respect to $r$, we get
\begin{equation}\label{drhodr}
	8\pi\frac{d\rho}{dr}=-\frac{2r(K-1)}{R^{4}}\frac{5K+\alpha+K(K-\alpha)\frac{r^{2}}{R^{2}}}{\left(1+K\frac{r^{2}}{R^{2}} \right)^{3}}.
\end{equation}
It is observed that $\frac{d\rho}{dr}(r=0)=0$ and $\frac{d\rho}{dr}(r=R)<0$ leads to the inequality
\begin{equation}\label{In2}
	K^{2}-K(\alpha-5)+\alpha \geq 0.
\end{equation}
The inequality (\ref{In2}) together with the condition $K>1$ give a bound for $\alpha$ as
\begin{equation}\label{In3b}
	0\leq\alpha<\frac{K(K+5)}{K-1}.
\end{equation}
The expression for $p_{\perp}$ is
\begin{equation}\label{pp3}
	8\pi p_{\perp}=\frac{4K-4+X_{1}\frac{r^{2}}{R^{2}}+X_{2}\frac{r^{4}}{R^{4}}+X_{3}\frac{r^{6}}{R^{6}}}{R^2\left(4+Y_{1}\frac{r^{2}}{R^{2}}+Y_{2}\frac{r^{4}}{R^{4}}+Y_{3}\frac{r^{6}}{R^{6}}+4K^{3}\frac{r^{8}}{R^{8}} \right)},
\end{equation}
where, $X_{1}=4K^{2}+(-12\alpha-16)K+12\alpha+12$, $X_{2}=6K^{3}+(-10\alpha-22)K^2+(4\alpha+14)K+6\alpha+2$,
$X_{3}=K^4+(-2\alpha-4)K^3+(\alpha^{2}+2\alpha+6)K^2+(-2\alpha^{2}-2\alpha-4)K+\alpha^{2}+2\alpha+1$, $Y_{1}=12K+4$, $Y_{2}=12K^{2}+12K$
and $Y_{3}=4K^{3}+12K^2$.

The condition $p_{\perp}>0$ at the boundary $ r = R $ imposes a restriction on $ K $ and $\alpha$ respectively given by 
\begin{equation}\label{In3a}
	K > 2 \sqrt{3}-1
\end{equation}
and 
\begin{equation}\label{In3}
	0\leq\alpha<\frac{10+5K+K^{2}}{K-1}-\sqrt{\frac{89+102K+57K^2+8K^3}{\left(K-1 \right)^{2}}}.
\end{equation}
The expression for $\frac{dp_{\perp}}{dr}$ is given by
\begin{equation}\label{dppdr}
	\frac{dp_{\perp}}{dr}=\frac{-r\left(8K^2+(12\alpha+8)K-12\alpha-16+A_{1}\frac{r^{2}}{R^{2}}+A_{2}\frac{r^{4}}{R^{4}}+A_{3}\frac{r^{6}}{R^{6}}+A_{4}\frac{r^{8}}{R^{8}} \right)}{R^{4}\left(2+B_{1}\frac{r^{2}}{R^{2}}+B_{2}\frac{r^{4}}{R^{4}}+B_{3}\frac{r^{6}}{R^{6}}+B_{4}\frac{r^{8}}{R^{8}}+B_{5}\frac{r^{10}}{R^{10}}+2K^4\frac{r^{12}}{R^{12}} \right)},
\end{equation}
where, $A_{1}=-4K^3+(28-4\alpha)K^2+(16\alpha-20)K-12\alpha-4$, $A_{2}=3K^4+(-4\alpha-4)K^3+(-3\alpha^{2}-28\alpha-30)K^2+(6\alpha^{2}+44\alpha+36)K-3\alpha^{2}-12\alpha-5$,
$A_{3}=10K^4+(-16\alpha-36)K^{3}+(-2\alpha^{2}+4\alpha+16)K^2+(4\alpha^{2}+16\alpha+12)K-2\alpha^{2}-4\alpha-2$, 
$A_{4}=K^{5}+(-2\alpha-4)K^{4}+(\alpha^{2}+2\alpha+6)K^{3}+(-2\alpha^{2}-2\alpha-4)k^{2}+\alpha^{2}+2\alpha+1$, $B_{1}=8k+4$,
$B_{2}=12K^{2}+16K+2$, $B_{3}=8K^{3}+24K^{2}+8K$, $2K^4+16K^3+12K^2$ and $B_{4}=4K^{4}+8K^{3}$. 

The value of $\frac{dp_{\perp}}{dr}=0$ at the origin
and $\frac{dp_{\perp}}{dr}(r=R)<0$ gives the following bounds for $ K $ and $\alpha$ respectively
\begin{equation}\label{In4a}
	2 \sqrt{13}-5 < K < 5
\end{equation}
and
\begin{equation}\label{In4}
	0 \leq \alpha < \frac{K^3+10 K^2+25 K-20}{K^2-6 K+5}+\sqrt{\frac{16 K^5+233 K^4+252 K^3+278 K^2-788 K+265}{\left(K^2-6 K+5\right)^2}}
\end{equation}

In order to examine the strong energy condition, we evaluate the expression $\rho-p_{r}-2p_{\perp}$ at the centre and on the boundary
of the star. It is found that
\begin{equation}\label{In5}
	\left(\rho-p_{r}-2p_{\perp} \right)(r=0)=0,
\end{equation}
and $\left(\rho-p_{r}-2p_{\perp} \right)(r=R)>0$ gives the bound on $ K $ and $\alpha$, namely
\begin{equation}\label{In6a}
	1< K < 1+2 \sqrt{6}
\end{equation}
\begin{equation}\label{In6}
	0\leq\alpha<\frac{8+3K+K^2}{K-1}+\sqrt{\frac{41+46K+49K^2+8K^3}{\left(K-1 \right)^{2}}}.
\end{equation}

The expressions for adiabatic sound speed $\frac{dp_{r}}{d\rho}$ and $\frac{dp_{\perp}}{d\rho}$ in the radial and transverse directions, respectively, are given by 
\begin{equation}\label{dprdrho}
	\frac{dp_{r}}{d\rho}=\frac{1+2K-K\frac{r^{2}}{R^{2}}}{5k+\alpha+K(K-\alpha)\frac{r^{2}}{R^{2}}} ,
\end{equation}
and
\begin{equation}\label{dppdrho}
	\frac{dp_{\perp}}{d\rho}=\frac{\left(1+K\frac{r^{2}}{R^{2}} \right)^{3}\left[8K^{2}+(12\alpha+8)K-12\alpha-16+C_{1}\frac{r^{2}}{R^{2}}+C_{2}\frac{r^{4}}{R^{4}}+C_{3}\frac{r^{6}}{R^{6}}+C_{4}\frac{r^{8}}{R^{8}} \right]}{2(K-1)\left[5K+\alpha+K(K-\alpha)\frac{r^{2}}{R^{2}} \right]\left[2+D_{1}\frac{r^{2}}{R^{2}}+D_{2}\frac{r^{4}}{R^{4}}+D_{3}\frac{r^{6}}{R^{6}}+D_{4}\frac{r^{8}}{R^{8}}+D_{5}\frac{r^{10}}{R^{10}}+2K^4\frac{r^{12}}{R^{12}} \right]},
\end{equation}
where, $C_{1}=-4K^{3}+(18-4\alpha)K^2+(16\alpha-20)K-12\alpha-4$, $C_{2}=3K^{4}+(-4\alpha-4)K^{3}+\left(-3\alpha^{2}-28\alpha-30 \right)K^{2}+\left(6\alpha^{2}+44\alpha+36 \right)K-3\alpha^{2}-12\alpha-5$,
$C_{3}=10K^{4}+(-16\alpha-36)K^{3}+\left(-2\alpha^{2}+4\alpha+16 \right)K^2+\left(4\alpha^{2}+16\alpha+12 \right)K-2\alpha^{2}-4\alpha-2$,
$C_{4}=K^{5}+(-2\alpha-4)K^{4}+\left(\alpha^{2}+2\alpha+6 \right)K^{3}+\left(-2\alpha^{2}-2\alpha-4 \right)K^{2}+\left(\alpha^{2}+2\alpha+1 \right)K$,
$D_{1}=8K+4$, $D_{2}=12K^{2}+16K+2$, $D_{3}=8K^{3}+24K^{2}+8K$, $D_{4}=2K^{4}+16K^{3}+12K^{2}$ and $D_{5}=4K^{4}+8K^{3}$.

The condition $ 0 \leq \frac{dp_{r}}{d\rho}\leq 1$  is evidently satisfied at the centre whereas at the boundary it gives a restriction on $ \alpha $ as
\begin{equation}\label{Inpre7}
	0\leq\alpha < \frac{K^2+4 K-1}{K-1} , ~K > 1.
\end{equation}
Further $\frac{dp_{\perp}}{d\rho}\leq 1$ at the centre will lead to the following inequalities
\begin{equation}\label{Inprepre7}
	K > \frac{4}{3}
\end{equation}
and
\begin{equation}\label{Inpreprepre7}
	0\leq \alpha <\frac{1}{2} (3 K-4).
\end{equation}
Moreover at the boundary $(r=R)$, we have the following restrictions on $ K $ and $ \alpha $.

\begin{equation}\label{Inpreprepreprepre7}
	-5 + 2 \sqrt{13} \leq K < 5
\end{equation}
and 
\begin{equation}\label{Inprepreprepreprepre7}
	0\leq \alpha \leq \frac{K^3+10 K^2+25 K-20}{K^2-6 K+5}+\sqrt{\frac{16 K^5+233 K^4+252 K^3+278 K^2-788 K+265}{\left(K^2-6 K+5\right)^2}},
\end{equation}

The necessary condition for the model to represent a stable relativistic star is that $\Gamma>\frac{4}{3}$ throughout the star. $\Gamma>\frac{4}{3}$ 
at $r=0$ gives a bound on $\alpha$ which is identical to (\ref{In1}). Further, $\Gamma\to\infty$ as $r\to R$ and hence the condition is automatically satisfied. It can be noticed that $E=0$ at $r=0$, showing the regularity of the charged 
distribution. \\

The upper limits of $ \alpha $ in the inequalities (\ref{In1}), (\ref{In3b}), (\ref{In3}), (\ref{In4}), (\ref{In6}), (\ref{Inpre7}) and (\ref{Inpreprepre7}) for different permissible values of $ K $ are shown in Table~\ref{tab:1}. It can be noticed that for $ 2.4641 < K \leq 3.7641 $ the bound for $ \alpha $ is $ 0 \leq \alpha \leq 0.6045. $

\pagebreak

\begin{table}[hbtp]
\caption{The upper limits of $ \alpha $ for different permissible values of $ K $.}
\label{tab:1}
\begin{tabular}{cccccccc}
\toprule
\multicolumn{1}{c}{}&\multicolumn{7}{c}{Inequality Numbers} \\ \cline{2-8}
$ K $ & (\ref{In1}) & (\ref{In3b}) & (\ref{In3}) & (\ref{Inpre7}) & (\ref{Inpreprepre7}) & (\ref{In4}) & (\ref{In6}) \\ \hline
2.4641 & 5.4641 & 12.5622 & \textbf{0.0000} & 10.1962 & 1.6962 & 0.0802 & 30.9893 \\ 
2.5041 & 5.5041 & 12.4932 & \textbf{0.0170} & 10.1635 & 1.7562 & 0.0938 & 30.6186 \\ 
2.6041 & 5.6041 & 12.3445 & \textbf{0.0599} & 10.0977 & 1.9062 & 0.1287 & 29.7861 \\ 
2.7041 & 5.7041 & 12.2250 & \textbf{0.1036} & 10.0514 & 2.0562 & 0.1648 & 29.0693 \\ 
2.8041 & 5.8041 & 12.1299 & \textbf{0.1480} & 10.0213 & 2.2062 & 0.2021 & 28.4488 \\ 
2.9041 & 5.9041 & 12.0552 & \textbf{0.1931} & 10.0048 & 2.3562 & 0.2405 & 27.9094 \\
3.0041 & 6.0041 & 11.9980 & \textbf{0.2388} & 10.0000 & 2.5062 & 0.2798 & 27.4388 \\
3.1041 & 6.1041 & 11.9557 & \textbf{0.2852} & 10.0052 & 2.6562 & 0.3201 & 27.0271 \\
3.2041 & 6.2041 & 11.9263 & \textbf{0.3321} & 10.0189 & 2.8062 & 0.3612 & 26.6662 \\ 
3.3041 & 6.3041 & 11.9082 & \textbf{0.3795} & 10.0401 & 2.9562 & 0.4030 & 26.3495 \\
3.4041 & 6.4041 & 11.8998 & \textbf{0.4275} & 10.0679 & 3.1062 & 0.4457 & 26.0714 \\
3.5041 & 6.5041 & 11.9002 & \textbf{0.4760} & 10.1015 & 3.2562 & 0.4890 & 25.8272 \\
3.6041 & 6.6041 & 11.9082 & \textbf{0.5251} & 10.1401 & 3.4062 & 0.5330 & 25.6130 \\
3.7041 & 6.7041 & 11.9230 & \textbf{0.5745} & 10.1833 & 3.5562 & 0.5776 & 25.4254 \\ 
3.7541 & 6.7541 & 11.9327 & \textbf{0.5995} & 10.2065 & 3.6312 & 0.6001 & 25.3407 \\ 
3.7641 & 6.7641 & 11.9348 & \textbf{0.6045} & 10.2112 & 3.6462 & 0.6047 & 25.3244 \\ \hline
\end{tabular}
\end{table}

\section{Application to Compact Stars and Discussion}
\label{sec:4}

In order to compare the charged anisotropic model on pseudo-spheroidal spacetime with observational data, we have considered the
pulsar PSR J1614-2230 whose estimated mass and radius are $1.97M_{\odot}$ and $9.69\; km$. On substituting these values in equation
(\ref{M}) we have obtained the values of adjustable parameters $K$ and $\alpha$ as $K=3.58524$ and $\alpha=0.292156$ respectively
which are well inside their permitted limits. Similarly assuming the estimated masses and radii of some well known pulsars like 
4U 1820-30, PSR J1903+327, 4U 1608-52, Vela X-1, PSR J1614-2230, Cen X-3, we have displayed the values of the parameters
$K$ and $\alpha$, the central density $\rho_{c}$, surface density $\rho_{R}$, the compactification factor $u=\frac{M}{R}$,
$\frac{dp_{r}}{d\rho}(r=0)$ and charge $ Q $ inside the star in Table~\ref{tab:3}. From the table it is clear that our model is in good agreement with the 
most recent observational data of pulsars given by \cite{Gangopadhyay13}. \\

\begin{table}[h]
\caption{Estimated physical values based on the observational data}
\label{tab:3}
\begin{tabular}{lllllllll}
\hline\noalign{\smallskip}
\textbf{STAR} & $\mathbf{K} $ & {$ \mathbf{M} $} & {$ \mathbf{R} $} & {$ \mathbf{\rho_c} $} & {$ \mathbf{\rho_R} $} & {$ \mathbf{u (=\frac{M}{R})} $} & $ \mathbf{\left(\frac{dp_r}{d \rho}\right)_{r=0}} $ & $ \mathbf{Q} $ \\
& & $ \mathbf{(M_\odot)} $ & $ \mathbf{(Km)} $ & \textbf{(MeV fm{$\mathbf{^{-3}}$})} & \textbf{(MeV fm{$\mathbf{^{-3}}$})} & & & $ \mathbf{Coulomb} $ \\
\noalign{\smallskip}\hline\noalign{\smallskip}
\textbf{4U 1820-30} 	  & 2.815 & 1.58  & 9.1   & 1980.14 & 250.46 & 0.256 & 0.461 &  $ 4.031\times10^{20} $\\
\textbf{PSR J1903+327} 	  & 2.880 & 1.667 & 9.438 & 1906.90 & 235.92 & 0.261 & 0.460 &  $ 4.184\times10^{20} $\\
\textbf{4U 1608-52} 	  & 3.122 & 1.74  & 9.31  & 2212.22 & 252.97 & 0.276 & 0.455 &  $ 4.127\times10^{20} $\\
\textbf{Vela X-1} 	  & 3.078 & 1.77  & 9.56  & 2054.99 & 238.25 & 0.273 & 0.456 &  $ 4.240\times10^{20} $\\
\textbf{PSR J1614-2230}   & \textbf{3.585} & \textbf{1.97} & \textbf{9.69} & \textbf{2487.35} & \textbf{248.17} & \textbf{0.300} & \textbf{0.448} & $ \mathbf{4.262\times10^{20}} $ \\
\textbf{Cen X-3}          & 2.589 & 1.49  & 9.178 & 1705.08 & 233.65 & 0.239 & 0.466 & $ 4.044\times10^{20} $ \\
\noalign{\smallskip}\hline
\end{tabular} 
\end{table}

\begin{figure}
\includegraphics[scale = 1.25]{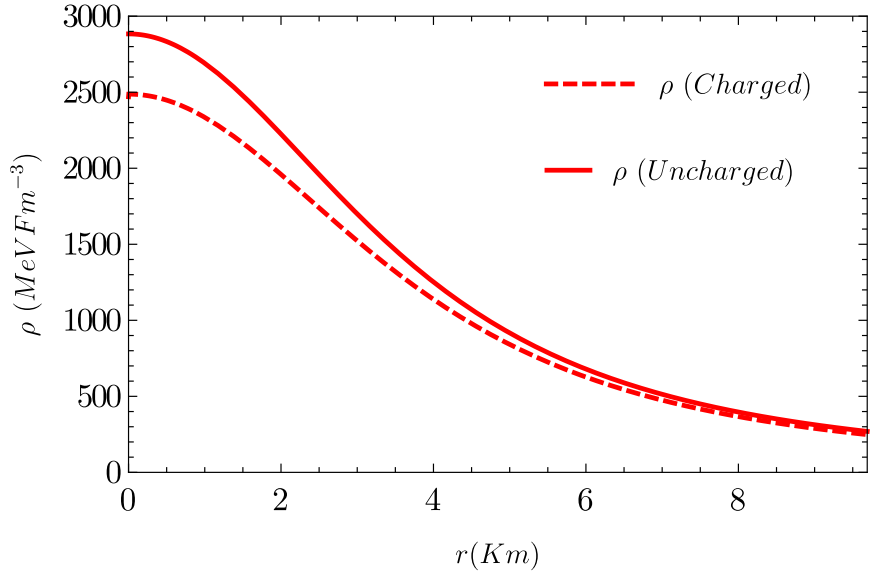} 
\caption{Variation of density against radial variable $r$. 
\label{fig:1}}
\end{figure}

\begin{figure}
\includegraphics[scale = 1.25]{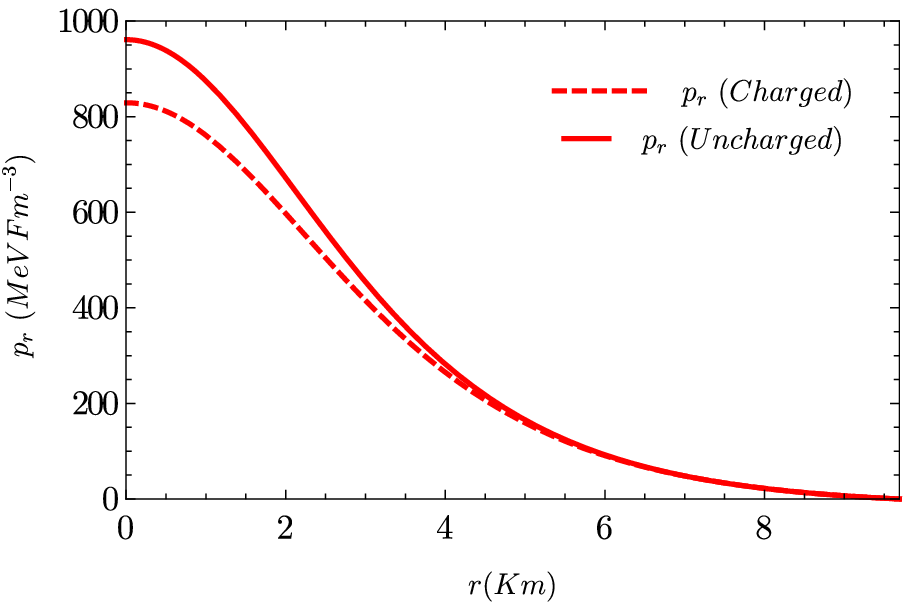}
\caption{Variation of radial pressures against radial variable $r$.
\label{fig:2}}
\end{figure}

\begin{figure}
\includegraphics[scale = 1.25]{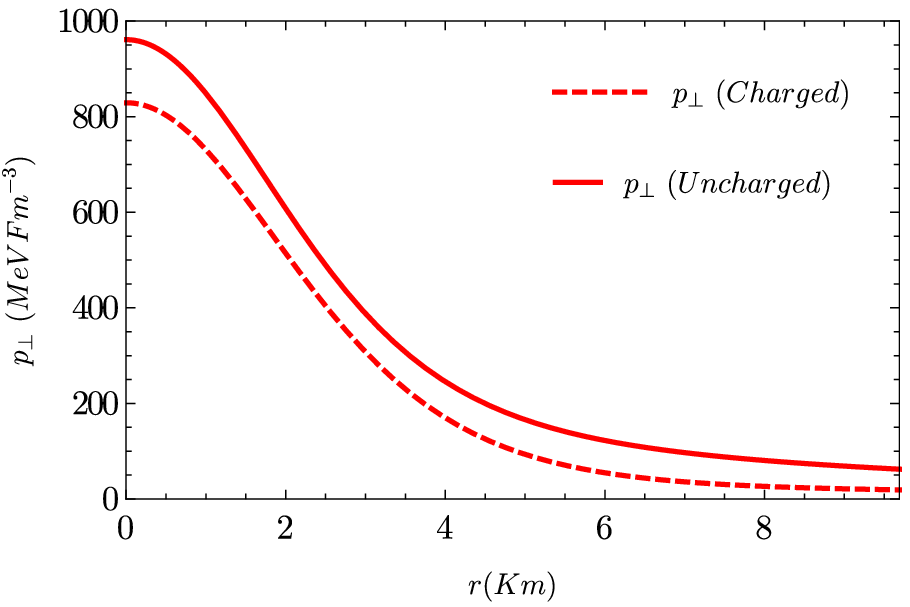}
\caption{Variation of transverse pressures against radial variable $r$ 
\label{fig:3}}
\end{figure}

In order to examine the nature of physical quantities throughout the distribution, we have considered a particular star PSR J1614-2230, whose tabulated mass and radius are 
$M=1.97M_{\odot},\;R=9.69\;km$. Choosing $K=3.58524$ and $\alpha=0.292156$, we have shown the variations of density and pressures in
both the charged and uncharged cases in Figure~\ref{fig:1}, Figure~\ref{fig:2} and Figure~\ref{fig:3}. It can be noticed that the 
pressure is decreasing radially outward. The density in the uncharged case is always greater than the density in the charged case.
Similarly the radial pressure $p_{r}$ and transverse pressure $p_{\perp}$ are decreasing radially outward. Similar to that of 
density, $p_{r}$ and $p_{\perp}$ in the uncharged case accommodate more values compared to charged case.

\begin{figure}
\includegraphics[scale = 1.25]{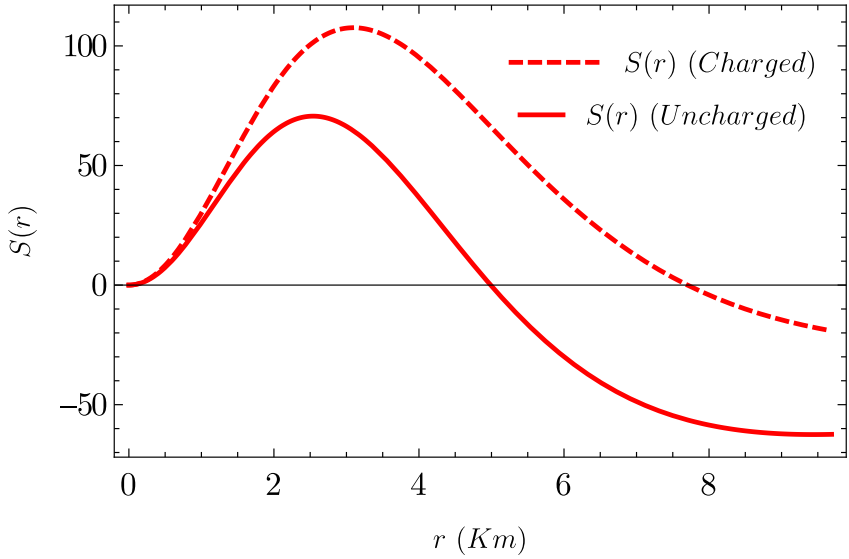}
\caption{Variation of anisotropies against radial variable $r$. 
\label{fig:4}}
\end{figure}

\begin{figure}
\includegraphics[scale = 1.25]{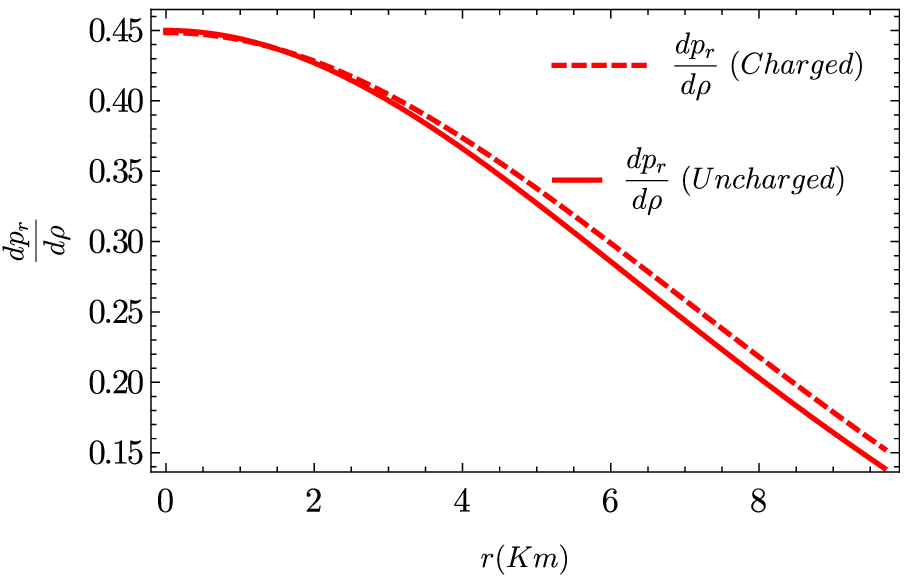}
\caption{Variation of $ \frac{1}{c^2}\frac{dp_r}{d\rho} $ against radial variable $r$. 
\label{fig:5}}
\end{figure}

\begin{figure}
\includegraphics[scale = 1.25]{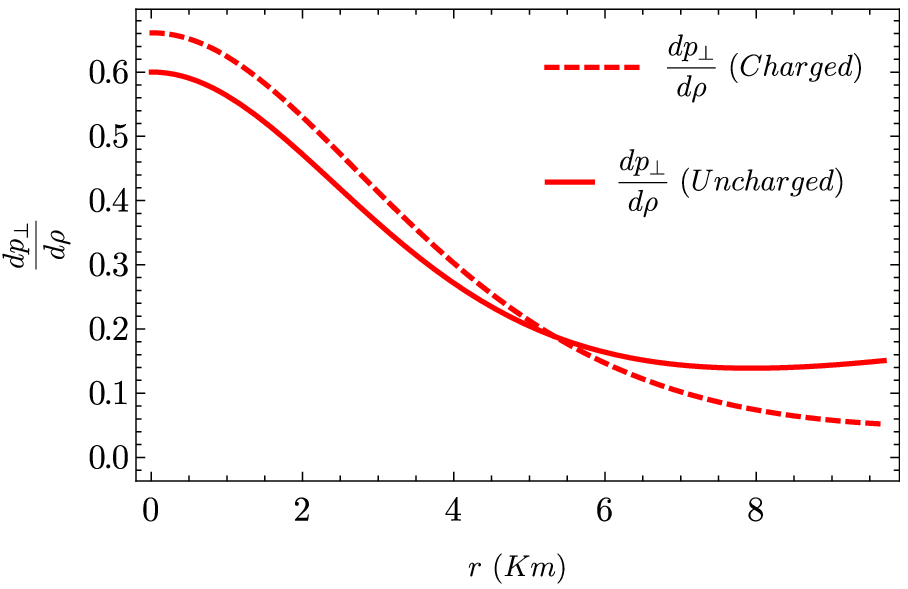}
\caption{Variation of $ \frac{1}{c^2}\frac{dp_\perp}{d\rho} $ against radial variable $r$.
\label{fig:6}}
\end{figure}

\begin{figure}
\includegraphics[scale = 1.25]{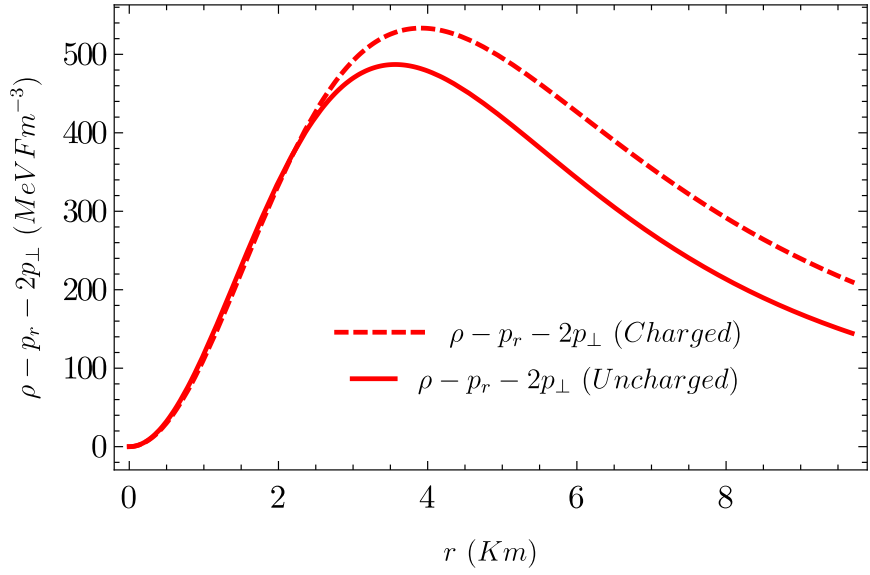}
\caption{Variation of strong energy condition against radial variable $r$. 
\label{fig:7}}
\end{figure}

The variation of anisotropy shown in Figure~\ref{fig:4} is initially decreasing with negative values reaches a minimum and then increases.
In this case also anisotropy takes lesser values in the charged case compared to uncharged case. The square of sound in the radial
and transverse direction (i.e. $\frac{dp_{r}}{d\rho}$ and $\frac{dp_{\perp}}{d\rho}$) are shown in Figure~\ref{fig:5} and 
Figure~\ref{fig:6} respectively and found that they are less than 1. The graph of $\rho-p_{r}-2p_{\perp}$ against radius is plotted
Figure~\ref{fig:7}. It can be observed that it is non-negative for $ 0 \leq r \leq R $ and hence strong energy condition is satisfied throughout the star. \\

\begin{figure}
\includegraphics[scale = 0.9]{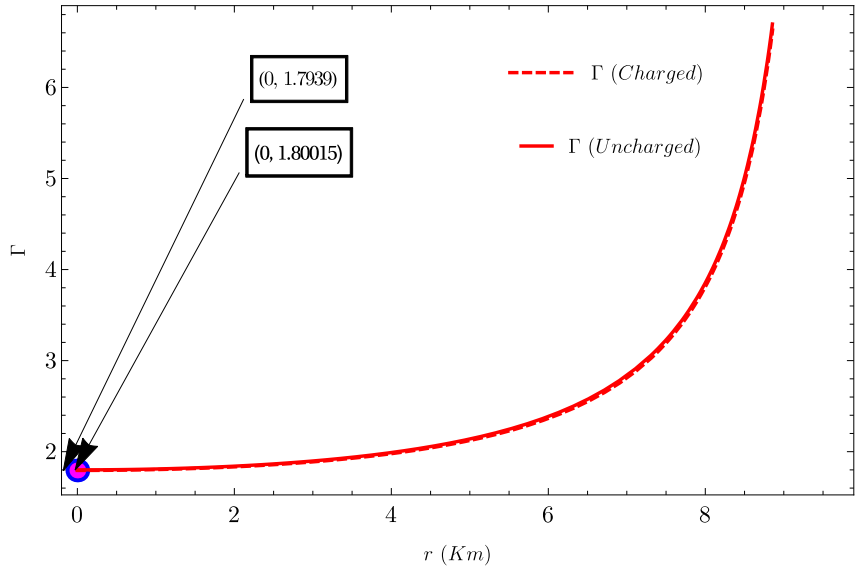}
\caption{Variation of $ \Gamma $ against radial variable $r$. 
\label{fig:8}}
\end{figure}

\begin{figure}
\includegraphics[scale = 1.25]{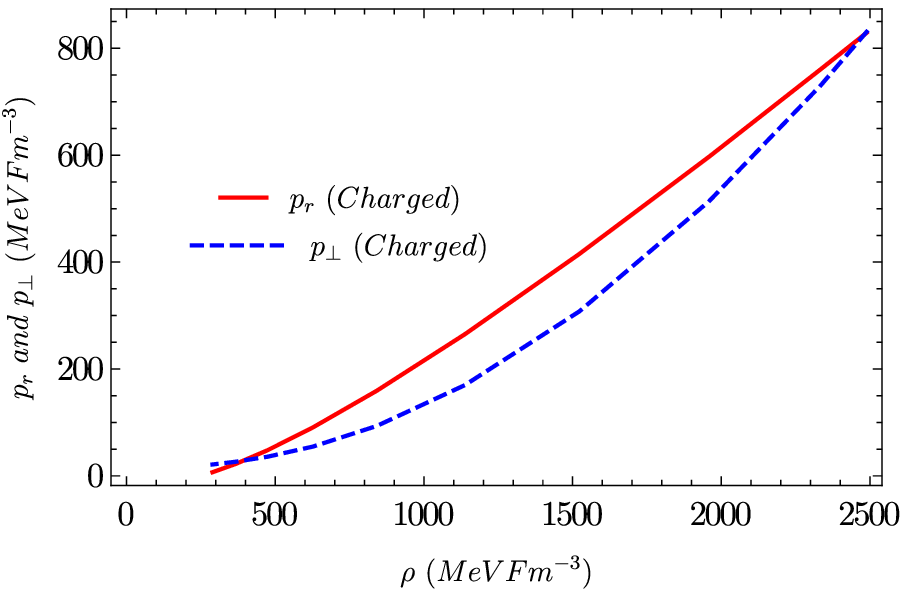}
\caption{Variation of pressures against density for charged case. 
\label{fig:9}}
\end{figure}

\begin{figure}
\includegraphics[scale = 1.25]{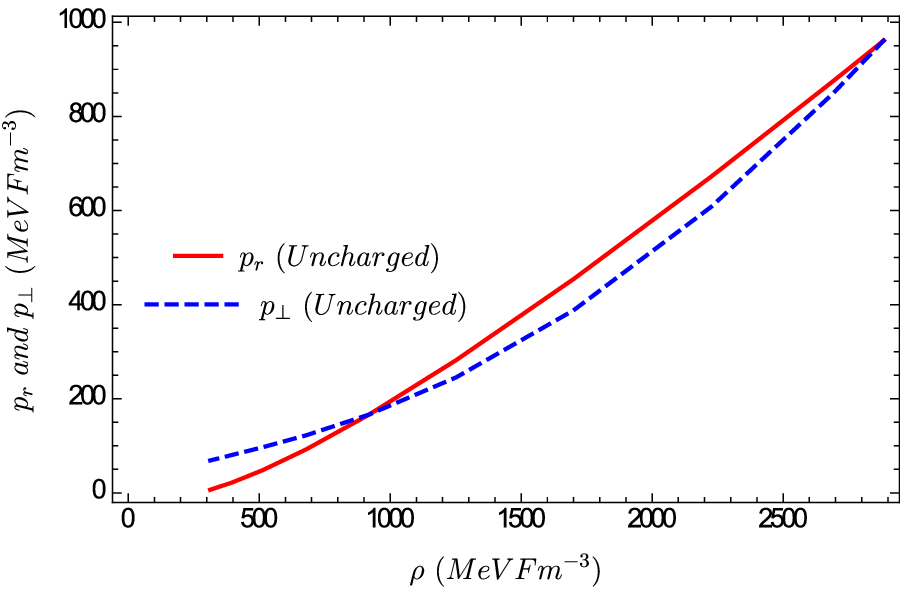}
\caption{Variation of pressures against density for uncharged case. 
\label{fig:10}}
\end{figure}

\begin{figure}
\includegraphics[scale = 1.25]{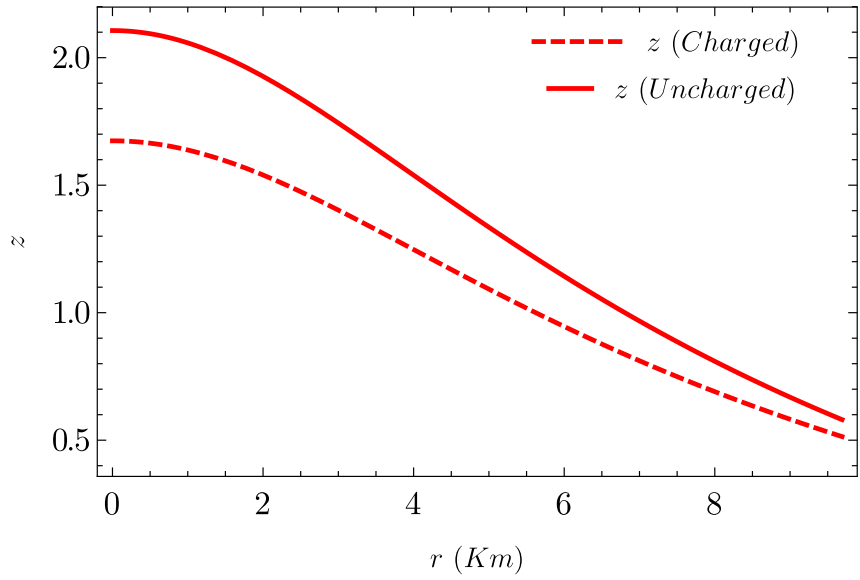}
\caption{Variation of gravitational redshift against radial variable $r$. 
\label{fig:11}}
\end{figure}

\begin{figure}
\includegraphics[scale = 1.25]{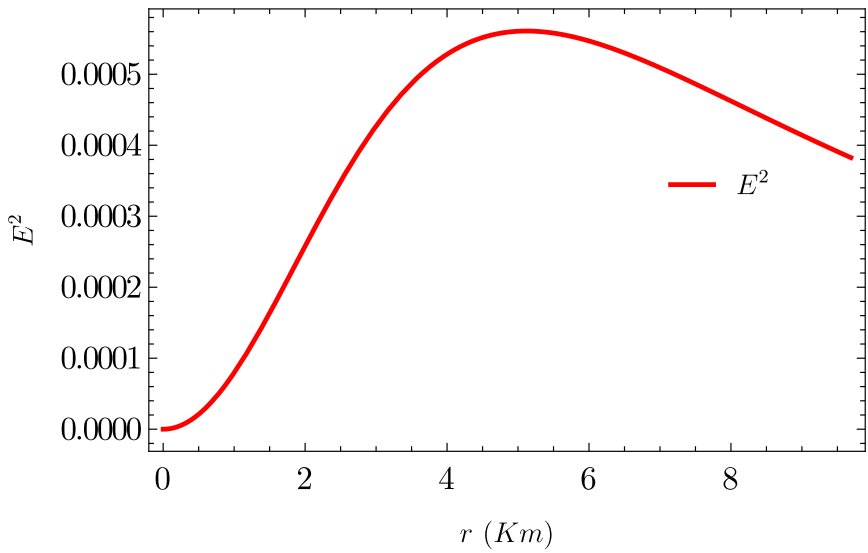}
\caption{Variation of $ E^2 $ against radial variable $r$. 
\label{fig:12}}
\end{figure}

A necessary condition for the exact solution to represent stable relativistic star is that the relativistic adiabatic index given by $ \Gamma = \frac{\rho + p_r}{p_r} \frac{d p_r}{d \rho} $ should be greater than $ \frac{4}{3}. $ The variation of adiabatic index throughout the star is shown in Figure~\ref{fig:8} and it is found that $ \Gamma > \frac{4}{3} $ throughout the distribution both in charged and uncharged case. Though we have not assumed any equation of state in the explicit form $ p_r = p_r (\rho) $ and $ p_\perp = p_\perp (\rho) $, we have shown the relation between $ p_r , p_\perp $ against $ \rho $ in the graphical form as displayed in  Figure~\ref{fig:9} and Figure~\ref{fig:10}. For a physically acceptable relativistic star the gravitational redshift must be positive and finite at the centre and on the boundary. Further it should be a decreasing function of $ r $. Figure~\ref{fig:11} shows that this is indeed the case. Finally we have plotted the graph of $ E^2 $ against $ r $ which is displayed in Figure~\ref{fig:12}. Initially $ E^2 $ increases from $ 0 $ and reaches a maximum values and then decreases radially outward. The model reduces to the uncharged anisotropic distribution given by \cite{Ratanpal15} when $ \alpha = 0. $ 

\section*{Acknowledgement}
The authors would like to thank IUCAA, Pune for the facilities and hospitality provided to them for carrying out this work.

\pagebreak

\end{document}